\documentclass[a4paper,fleqn,usenatbib]{mnras}

\usepackage{newtxtext,newtxmath}
\usepackage[T1]{fontenc}
\usepackage{ae,aecompl}

\usepackage{graphicx}	% Including figure files
\usepackage{amssymb}	% Extra maths symbols
\usepackage{natbib}
\usepackage{soul}
\bibpunct{(}{)}{;}{a}{}{,}
\usepackage{graphicx}
\usepackage{multirow}
\usepackage{graphicx}
\usepackage{txfonts}
\usepackage{natbib}
\usepackage{xspace}
\usepackage{dcolumn}
\usepackage{longtable}
\usepackage{lscape}
\usepackage{soul}        

\usepackage{rotating}
\usepackage{afterpage}

\newcommand{\MSun}{M$_{\odot}$\xspace}
\newcommand{\MEarth}{M$_{\oplus}$\xspace}

\newcommand{\mic}{$\mu$m\xspace}
\newcommand{\as}{\hbox{$^{\prime\prime}$}\xspace}

    \title[Proxima Centauri's gravitational mass]{The gravitational mass of Proxima Centauri measured with SPHERE from a microlensing event.}

    \author[A. Zurlo et al.]{A. Zurlo$^{1,2,3}$\thanks{E-mail:
        alice.zurlo@mail.udp.cl}, R. Gratton$^{4}$, D. Mesa$^{5,4}$, S. Desidera$^{4}$, A. Enia$^{6}$, K. Sahu$^{7}$, J.-M. Almenara$^{8}$,\newauthor
      P. Kervella$^{9}$, H. Avenhaus$^{10}$, J. Girard$^{7}$,  M. Janson$^{11,12}$, E. Lagadec$^{13}$, M. Langlois$^{14,3}$, \newauthor
      J. Milli$^{15}$, C. Perrot$^{9}$, J.-E. Schlieder$^{16,11}$, C. Thalmann$^{10}$, A. Vigan$^{3}$, E. Giro$^{4}$, L. Gluck$^{17}$, \newauthor J. Ramos$^{11}$, A. Roux$^{17}$\\ \\
      $^{1}$N\'ucleo de Astronom\'ia, Facultad de Ingenier\'ia y Ciencias, Universidad Diego Portales, Av. Ejercito 441, Santiago, Chile\\
      $^{2}$Escuela de Ingenier\'ia Industrial, Facultad de Ingenier\'ia y Ciencias, Universidad Diego Portales, Av. Ejercito 441, Santiago, Chile \\
      $^{3}$Aix Marseille Universit\'e, CNRS, LAM - Laboratoire d'Astrophysique de Marseille, UMR 7326, 13388, Marseille, France  \\
      $^{4}$INAF-Osservatorio Astronomico di Padova, Vicolo dell'Osservatorio 5, Padova, Italy, 35122-I \\
      $^{5}$INCT, Universidad De Atacama, calle Copayapu 485, Copiap\'{o}, Atacama, Chile \\
      $^{6}$Dipartimento di Fisica e Astronomia Galileo Galilei, Universit\`a di Padova, Vicolo dell'Osservatorio 3, I-35122 Padova, Italy \\
      $^{7}$Space Telescope Science Institute, 3700 San Martin Drive, Baltimore, MD 21218, USA \\
      $^{8}$Observatoire de Gen\`eve, University of Geneva, 51 Chemin des Maillettes, 1290, Versoix, Switzerland\\  
      $^{9}$LESIA, Observatoire de Paris, CNRS, Universit\'e Paris Diderot, Universit\'e Pierre et Marie Curie, 5 place Jules Janssen, 92190 Meudon, France\\
      $^{10}$ETH Zurich, Institute for Astronomy, Wolfgang-Pauli-Strasse 27, 8093 Zurich, Switzerland\\  
      $^{11}$Max-Planck-Institut f\"ur Astronomie, K\"onigstuhl 17, 69117 Heidelberg, Germany\\
      $^{12}$Stockholm University, AlbaNova University Center, Stockholm, Sweden \\
      $^{13}$Universit\'e C\^ote d'Azur, Observatoire de la C\^ote d'Azur, CNRS, Lagrange, France \\
      $^{14}$CRAL, UMR 5574, CNRS, Universit\'e Lyon 1, 9 avenue Charles Andr\'e, 69561 Saint Genis Laval Cedex, France\\   
      $^{15}$European Southern Observatory, Alonso de Cordova 3107, Casilla 19, Vitacura, Santiago, Chile \\
      $^{16}$NASA Goddard Space Flight Center, Greenbelt, MD 20771, USA \\
      $^{17}$Universit\`e Grenoble Alpes, IPAG, F-38000 Grenoble, France  }

   \date{Accepted 2018 July 3. Received 2018 July 2; in original form 2018 February 5}

   \pubyear{2018}

\begin{document}
\label{firstpage}
\pagerange{\pageref{firstpage}--\pageref{lastpage}}
\maketitle

% Abstract of the paper
\begin{abstract}
Proxima Centauri, our closest stellar neighbour, is a low-mass M5 dwarf orbiting in a triple system. An Earth-mass planet with an 11 day period has been discovered around this star. The star's mass has been estimated only indirectly using a mass-luminosity relation, meaning that large uncertainties affect our knowledge of its properties. To refine the mass estimate, an independent method has been proposed: gravitational microlensing. By taking advantage of the close passage of Proxima Cen in front of two background stars, it is possible to { measure the astrometric shift caused by} the microlensing effect due to these close encounters and estimate the gravitational mass of the lens (Proxima Cen). Microlensing events occurred in 2014 and 2016 with impact parameters, the { closest approach of Proxima Cen to the background star}, of 1\farcs6 $\pm$ 0\farcs1 and 0\farcs5 $\pm$ 0\farcs1, respectively. Accurate measurements of the positions of the background stars during the last two years have been obtained with HST/WFC3, and with VLT/SPHERE from the ground. The SPHERE campaign started on March 2015, and continued for more than two years, covering 9 epochs. The parameters of Proxima Centauri's motion on the sky, along with the pixel scale, true North, and centering of the instrument detector were readjusted for each epoch using the background stars visible in the IRDIS field of view. The experiment has been successful and the { astrometric shift} caused by the microlensing effect has been measured for the { second event in 2016}. We used this measurement to derive a mass of 0.150$^{\textrm{+}0.062}_{-0.051}$ (an error of $\sim$ 40\%) \MSun for Proxima Centauri acting as a lens. This is the first and the only currently possible measurement of the gravitational mass of Proxima Centauri.    
\end{abstract}

% Select between one and six entries from the list of approved keywords.
% Don't make up new ones.

\section{Introduction}
\label{intro}

Proxima Centauri is the closest star to our Sun \citep[{ distance}=1.3 pc;][]{2014AJ....148...91L}. It is part of a triple system together with $\alpha$ Cen A and B \citep{2017A&A...598L...7K}. Proxima is a low-mass M5Ve, flaring dwarf star. It has very fast proper motion of -3775.75 mas/yr in right ascension and 765.54 mas/yr in declination \citep{2007A&A...474..653V}. \citet{2017A&A...598L...7K} recently demonstrated that Proxima is bound to $\alpha$ Cen AB with a high degree of confidence. A mass of $0.12\pm 0.02$~\MSun has been estimated using a mass-luminosity relation \citep[e.g.,][]{2015ApJ...804...64M}. Due to this method, this value suffers from significant uncertainties.  Especially for young systems, the masses estimated from astrometric binaries can be underestimated by some tens of percent (L. Rodet, private communication). The scatter on the measurements of astrometric and eclipsing binaries, which provide the error bars for the mass-luminosity derived mass, are higher for low-mass stars.

%The radius of Proxima has been measured with the interferometry technique by \citet{2003A&A...397L...5S}. 
Recently, a telluric planet with a minimum mass 1.3 \MEarth and an orbital period of 11.2 days, Proxima b, has been found in the habitable zone of Proxima Cen using the radial velocity method \citep{2016Natur.536..437A}. Moreover, \citet{2017ApJ...850L...6A} presented the discovery of warm dusts belts around the star. These discoveries brought even more interest to this star, and the need to determine its mass with higher accuracy. 

Very recently, \citet{2017Sci...356.1046S} for the first time astrometrically measured the gravitational deflection of starlight during a microlensing event to derive the mass of a star different from our Sun, the white dwarf Stein 2051 B. The fine measurement of the deflection using the { Wide Field Camera 3 (WFC3)} on the Hubble Space Telescope (HST), permitted a mass estimate of 0.675 $\pm$ 0.051 ($\pm$ 7.5\%) \MSun.  

\citet{2014ApJ...782...89S} predicted two microlensing events due to the close passage of Proxima Cen in front of two background stars, which they began monitoring using HST. Based on their first HST observations, they made precise predictions of the close encounters. The first passage would occur in 2014, when Proxima passed very close to a mag$_V$=20 background star (Source 1). The second microlensing event was predicted for the year 2016, due to close passage of another background star of mag$_V$ = 19.5, called by the authors Source 2. The closest approach by Proxima was expected to occur in February 2016, with an impact parameter of 0\farcs5 $\pm$ 0\farcs1. The expected { astrometric} shift of the centroid of Source 2 { on the sky} due to the microlensing effect had a maximum value of 1.5 mas.  Such an event was highlighted as a unique opportunity to estimate the mass of Proxima Cen. The full results of the HST program (PI: K. Sahu) will be presented in Sahu et al., in prep.

In this paper we present the astrometric monitoring of Proxima Cen and its close passage to Source 2 with the Spectro-Polarimetric High-contrast Exoplanet REsearch \citep[SPHERE;][]{2008SPIE.7014E..18B} instrument at the Very Large Telescope. SPHERE is a powerful high contrast instrument developed for the research of exoplanets and circumstellar discs. Its performance for high precision astrometry and photometry of companions was demonstrated during commissioning \citep[see, e.g.,][]{2016A&A...587A..55V, 2016A&A...587A..56M, 2016A&A...587A..57Z, 2016A&A...587A..58B}. SPHERE is composed of a common path and three scientific instruments: IRDIS \citep{dohlen2008}, a dual band imager, IFS \citep{Cl08}, an Integral Field Spectrograph, and ZIMPOL \citep{Th08}, a visible light polarimeter. SPHERE's powerful adaptive optics system \citep[i.e.,][]{Fusco:06}, and the relatively large { field of view (FoV)} of IRDIS for an { eXtreme Adaptive Optics (xAO)} fed instrument, allowed precise astrometric observations of the faint Proxima Cen from the ground. For the analysis presented in this paper, we used the near infrared arm of SPHERE to observe Proxima Centauri and the background stars using SPHERE from March 2015 to June 2017. This paper is structured as follows: in Sec.~\ref{s:obs} we present the observations and data reduction. In Sec.~\ref{s:irdifsdatared} we discuss the astrometric analysis, Sec.~\ref{s:dis} presents the discussion and conclusions are given in Sec.~\ref{s:conc}.     

%__________________________________________________________________
\section{Observations and data reduction}
\label{s:obs}
The observations of Proxima Centauri were carried out during the Guaranteed Time Observations (GTO) of SPHERE. { We acquired 9 epochs spanning over 2 years}. They are listed in Table~\ref{t:obs}. Most of these datasets (except the last two) were reduced and analysed by \citet{2017MNRAS.466L.118M}, where constraints on the mass of possible giant companions of Proxima b are presented. Nevertheless, we present here all of the observations for the sake of completeness. SPHERE was set in its IRDIFS mode, with the dual band imager IRDIS using the H2H3 dual-band filters \citep[H2=1.587 \mic and H3=1.667 \mic, { dual band imaging (DBI)} mode;][]{vigan2010} and the integral field spectrograph IFS covering the YJ band \citep[0.95--1.35 \mic; R=50;][]{zurlo2014}. Each dataset except the first one had a total exposure time of $\sim$ 1h, which allowed us to obtain deep high-contrast images of the region around Proxima Centauri within the IRDIS and IFS fields of view (FoVs). The size of the IRDIS FoV is 11\as $\times$ 11\as while IFS has a smaller FoV of 1.7\as $\times$ 1.7\as.

Initial IRDIS data processing was performed using the official pipeline of the consortium at the Data Center/IPAG in Grenoble \citep{2017arXiv171206948D}. IRDIS data, after the application of the appropriate calibrations (dark, flat and centering), were reduced with the {\it ad hoc} IDL routines presented in \citet{zurlo2014,2016A&A...587A..57Z}. The procedures apply the KLIP \citep{2012ApJ...755L..28S} principal component analysis (PCA) to attenuate the speckle noise and apply the angular differential imaging \citep[ADI;][]{2006ApJ...641..556M} technique to the IRDIS frames. For this particular analysis, we also produced median combined, derotated images cleaned with a spatial filter. { We first created a datacube which contained the frames of the observation derotated according to their position angles. A median filter that removes isolated high or low values was then applied and a median-collapsed image of the cube was produced.}

\begin{figure*}
%%%%%\vspace{8cm}
\begin{center}
\includegraphics[width=0.7\textwidth]{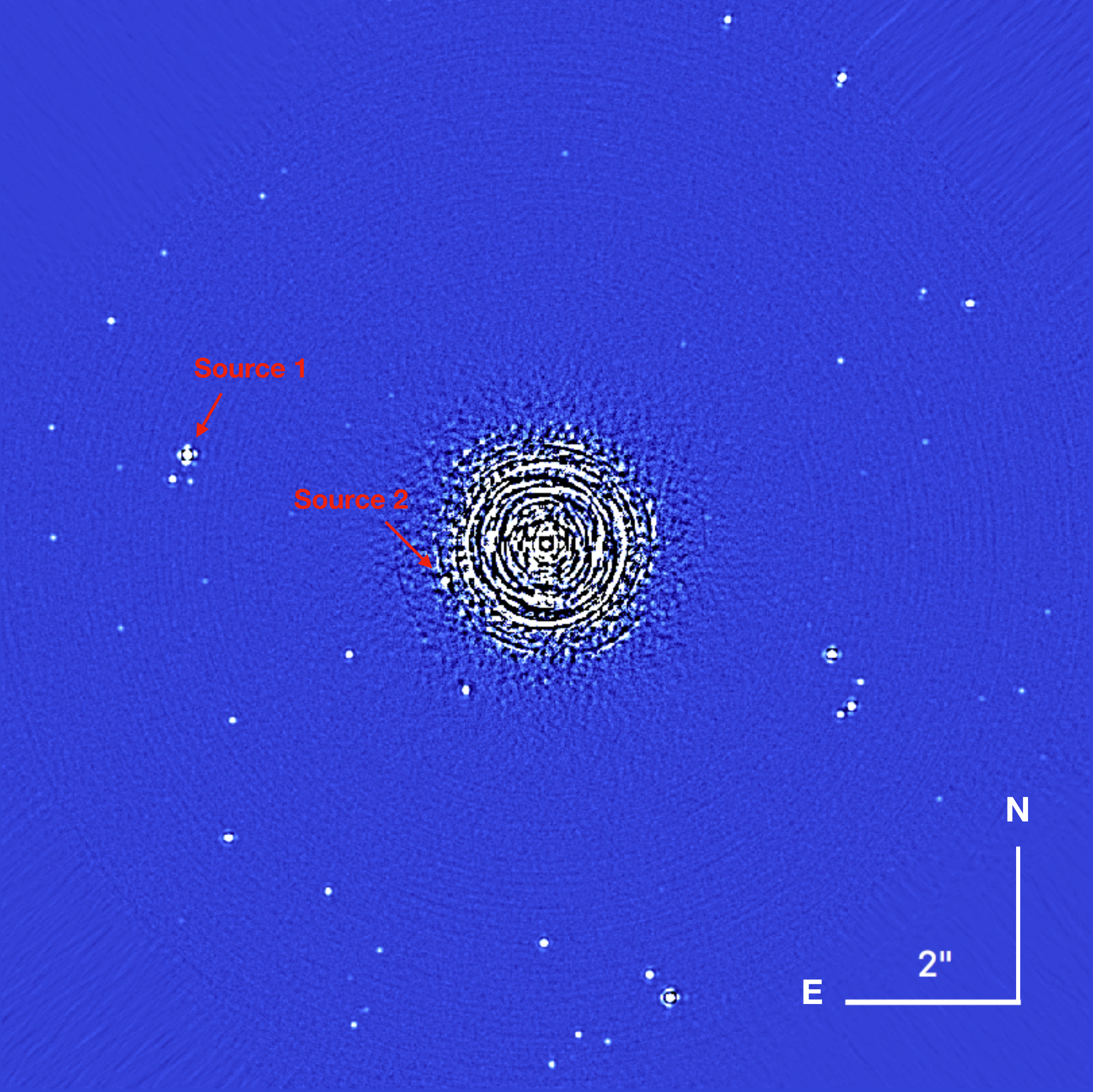}
\caption{IRDIS FoV for the April 2016 epoch. The image is derotated, median combined, and cleaned with a spatial filter. At the center of the image, inside the inner working angle (IWA), the speckle pattern dominates, in the outer part of the image our reduction method prevents the elongation of the stars' point spread functions (PSFs). The bars in the lower right provide the spatial scale. North is up and East is to the left. }
\label{f:irdis}
\end{center}
\end{figure*}

\begin{figure}
%%%%%\vspace{8cm}
\begin{center}
\includegraphics[width=0.5\textwidth]{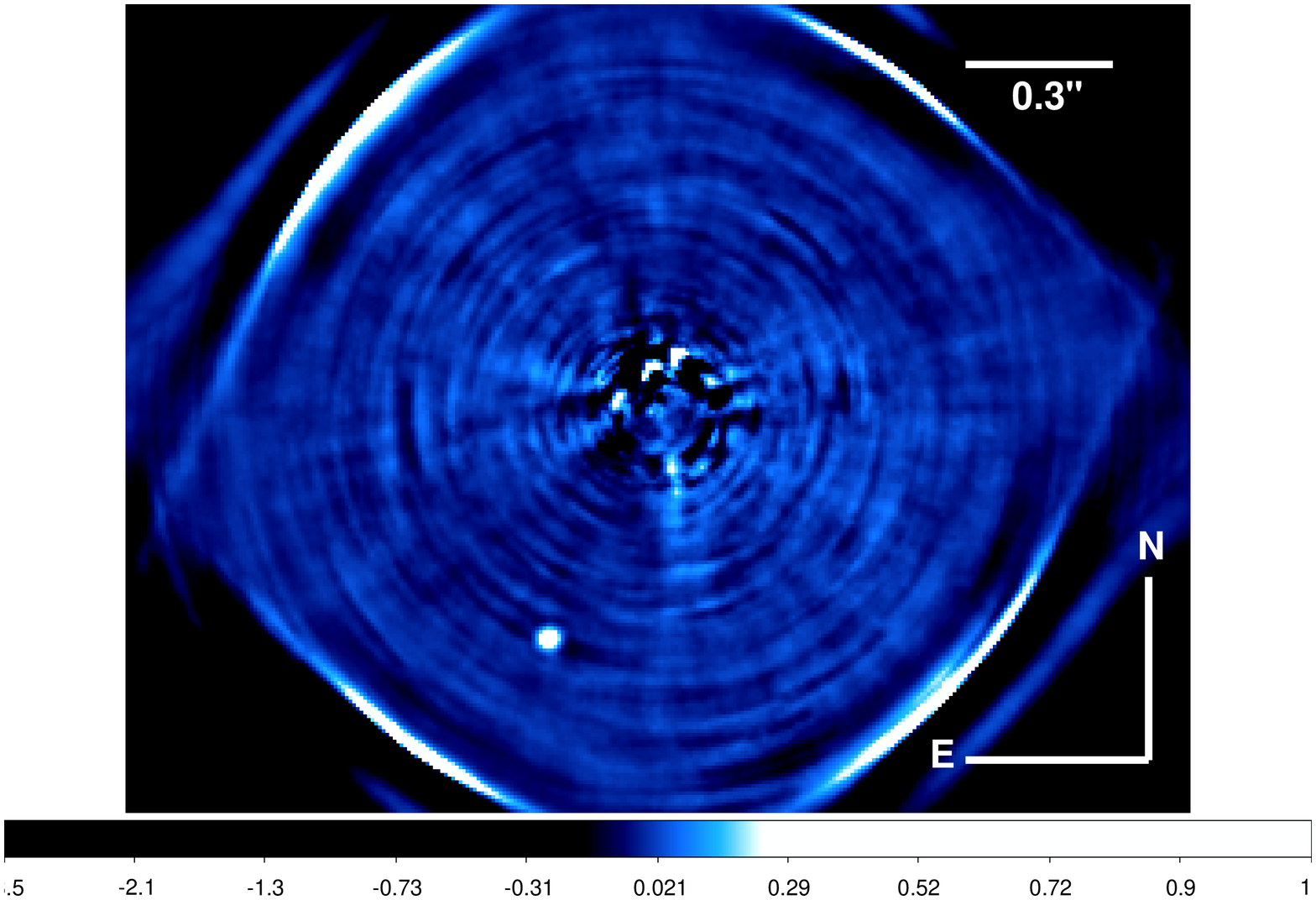}
\caption{IFS FoV for April 2016 epoch processed with the PCA method. Source 2 is the only star visible on the detector. The contrast of the Source 2 with respect to Proxima Cen is 11 mag in J band.}
\label{f:ifs}
\end{center}
\end{figure}

Following this approach, background stars are clearly visible inside the SPHERE adaptive optics outer working angle (OWA) and they do not suffer from elongation due to ADI self subtraction. The final image for one of the epochs is shown in Figure~\ref{f:irdis}. This median-combined method allowed an improvement on the determination of the astrometric positions of the stars outside the speckle pattern noise region very close to Proxima Cen. IFS data were reduced applying the algorithms presented in \citet{mesa2015}. We first applied the appropriate calibrations (dark, flat, spectral positions, wavelength calibration and instrument flat) using the standard DRH data reduction \citep{2008SPIE.7019E..39P}. The resulting datacubes were each comprised of 39 monochromatic frames. We then applied a custom PCA procedure to these datacubes. The reduced IFS image for the same epoch as shown for IRDIS in Figure~\ref{f:irdis} is presented in Figure~\ref{f:ifs}. The contrast reached for each observation is listed in Table 2 of \citet{2017MNRAS.466L.118M}.   

\begin{table*}
  \centering
 \begin{minipage}{140mm}
  \caption{List of SPHERE observations for Proxima Centauri. The { detector integration time} (DIT) represents the exposure time for each frame
expressed in seconds, nDIT represents the number of frames for each datacube of the dataset. The seeing was measured as the { full width at half maximum} (FWHM) with the DIMM instrument at ESO Paranal. The number of stars identified in the IRDIS { field of view} (FoV) for each epoch is listed in the last column.  \label{t:obs}}
  \begin{tabular}{ c c c c c c}
  \hline
    Date  &  nDIT;DIT(s) IRDIS  & nDIT;DIT(s) IFS & Rot.Ang. ($^{\circ}$) & DIMM Seeing (\as) & Stars Identified  \\
         \hline
         2015-03-30 &   3$\times$12;16  &  3$\times$12;16 &   3.12  &  0.93   & 23 \\
         2016-01-18  &  7$\times$40;16  &  7$\times$20;32 &  25.74  &  2.20   & 27 \\
         2016-02-17  &  11$\times$10;16 &  11$\times$5;32 &  13.52  &  1.86   & 32 \\
         2016-02-29  &  7$\times$10;16  &  7$\times$15;32 &  22.56  &  0.78   & 40 \\
         2016-03-27  &  5$\times$40;16  &  5$\times$25;32 &  25.69  &  2.08   & 44 \\
         2016-04-15  &  6$\times$40;16  &  6$\times$20;32 &  28.72  &  0.62   & 50 \\
         2016-07-20  & 5$\times$40;16 & 5$\times$20;32 &  21.14  & 0.40       & 38 \\
         2017-03-20  &  16$\times$2;64  &  16$\times$2;64 &  11.50  & 0.57    & 35 \\
         2017-06-15  & 5$\times$40;16 & 5$\times$20;32 & 21.26  & 1.69        & 35  \\
         
\hline
  \end{tabular}
\end{minipage}
\end{table*}

\section{IRDIFS astrometry}
\label{s:irdifsdatared}

\subsection{Measurement of the stellar positions}
\label{s:meas}

The total number of stars identified in in the IRDIS FoV in each epoch are listed in Table~\ref{t:obs}. To measure the position of each star, we employed three different IDL routines: \texttt{mpfit2dpeak} \citep{2009ASPC..411..251M}, \texttt{cntrd}, and \texttt{find}. For each star detected in the FoV, we identified which final image, the median combined image or the KLIP-ADI image, had a higher { signal-to-noise ratio (S/N)} and more symmetric { point spread function (PSF)} shape. For stars farther than 0\farcs8 from the central star we found in general that the median combined image was better because of the round shape of the PSFs. If the measurements of the three routines were consistent, the mean of the results was adopted, otherwise we excluded the routine(s) that did not converge. In most cases, all three routines gave consistent results. 

On the IFS final images we measured stellar positions by inserting negative-scaled PSF images { \citep[``fake negative planet'' technique;][]{2010Sci...329...57L, zurlo2014}} into the final image and shifting the simulated star position until the standard deviation was minimized in a small region around the PSF itself (0.8 $\lambda$/D). The median of the positions obtained when varying the number of principal components was assumed as the position of the object. Source 2 was detected with IFS for the epochs 2016-01-18, 2016-02-17, and 2016-02-29. For the epoch 2016-03-27 the star is partially visible on the edge of the IFS FoV. We did not consider this measurement in our analyses.    

\begin{figure*}
%%%%%\vspace{8cm}
\begin{center}
\includegraphics[width=\textwidth]{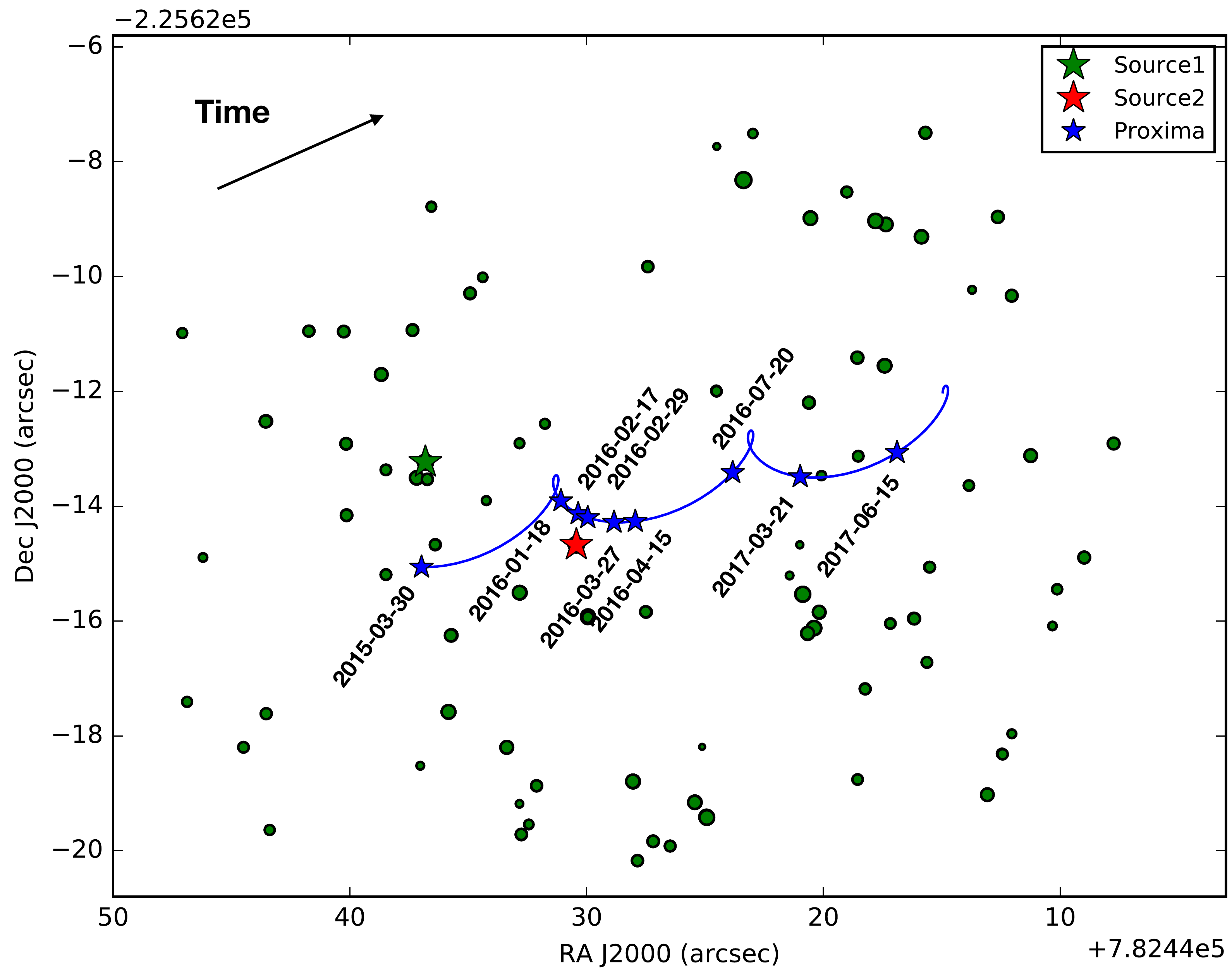}
\caption{Representation of the movement of Proxima (solid blue line) with respect to all the stars identified (green circles) during the 9 IRDIS epochs. { The movement of Proxima on the sky has been calculated using the nominal parameters of parallax and proper motions given by Hipparcos.} The size of the point is proportional to the logarithm of the flux of each star in H band. Proxima Cen is shown as a blue star for each epoch. Source 1 and Source 2 presented in \citet{2014ApJ...782...89S} are represented as a green and red star, respectively. The two stars close to Proxima Cen during the last two epochs are too faint to be detected during closest approach. They are $\sim$70 times fainter than Source 2. }
\label{f:pos}
\end{center}
\end{figure*}

\begin{figure*}
%%%%%\vspace{8cm}
\begin{center}
\includegraphics[width=0.6\textwidth]{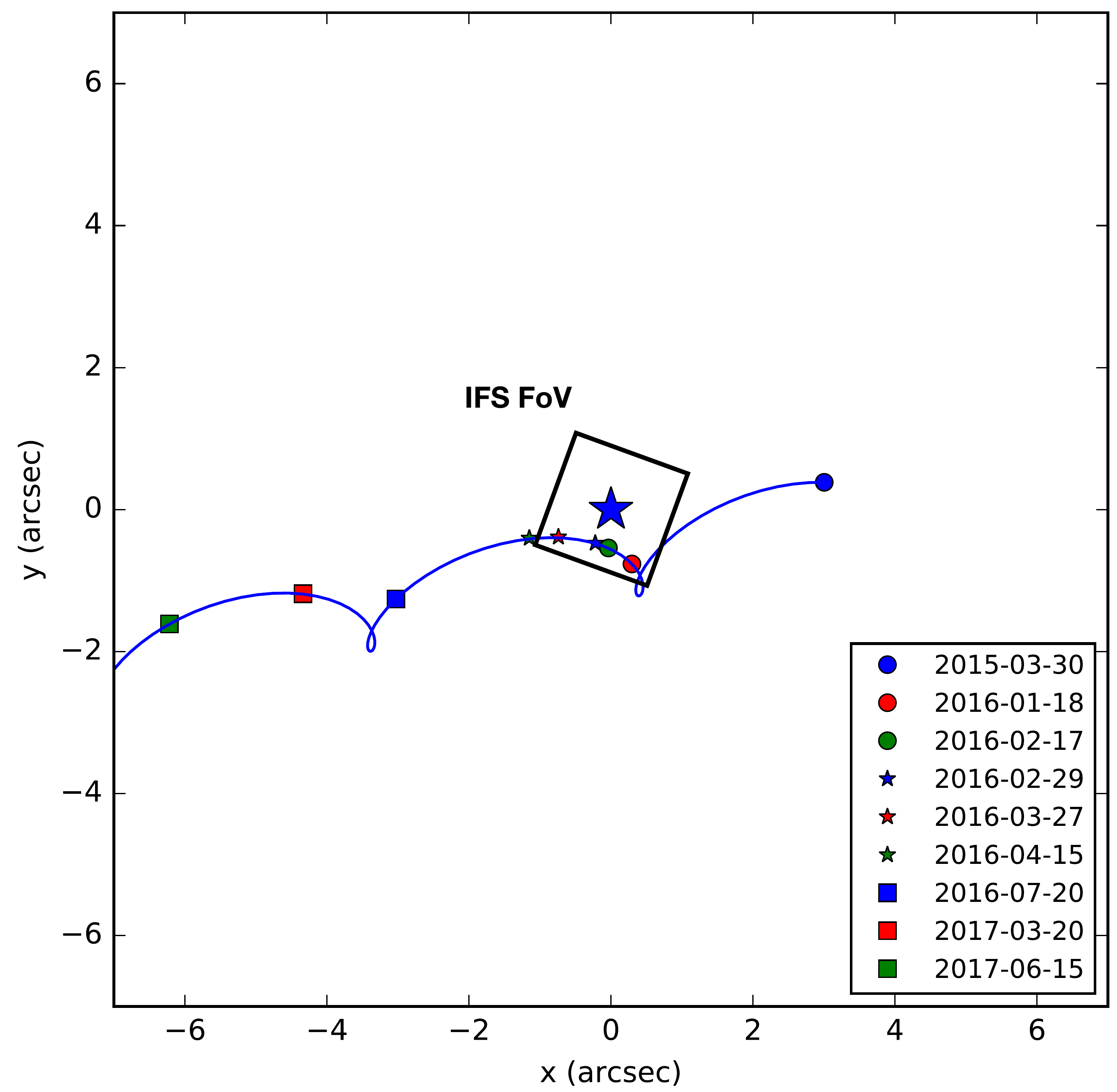}
\caption{Representation of the movement of Proxima Cen (solid blue line) reflected to Source 2. In the IRDIS FoV, Proxima appeared in the center of the image (blue star) and in each epoch the background star was identified at the positions represented in this plot (see legend).}
\label{f:pos_x}
\end{center}
\end{figure*}

The movement of Proxima Cen across the sky with respect to the other stars identified in the images for all epochs is presented in Figure~\ref{f:pos}. The relative displacement of Source 2 with respect to Proxima Cen as it is seen in the IRDIS detector for all epochs is shown in Figure~\ref{f:pos_x}. Source 2 was detected in all epochs. However, in the last observation (June 2017) it falls on the edge of the detector and is partially visible. 

%%%%%%%%%%%%%%%%%%%%%%%%%%%%%%%%%%%%%%%%%

%PICK UP HERE

%%%%%%%%%%%%%%%%%%%%%%%%%%%%%%%%%%%%%%%%%

\subsection{Accurate determination of the detector parameters and the movement Proxima of Cen }

After measuring the position of each star on the detector FoV for each epoch we ran a Markov Chain Monte Carlo (MCMC) to refine the detector parameters (position of the center, i.e. Proxima position, the value of the plate scale, and True North (TN)) for each epoch and the movement of Proxima on the sky. For these parameters the initial guesses of the MCMC were the standard IRDIS astrometric calibrations \citep[see][]{2016SPIE.9908E..34M} and the Hipparcos values for Proxima Cen's RA2000, DEC2000, $\mu_{RA}$, $\mu_{DEC}$, $\pi$. We estimated four detector parameters for each epoch ($x$ and $y$ coordinates of the center, plate scale and TN) and five parameters for the movement of Proxima (RA2000, DEC2000, $\mu_{RA}$, $\mu_{DEC}$, $\pi$) with this fit. This leads to 4$\times$9 + 5 = 41 parameters in total.

The positions of the background stars on the sky were calculated using the nominal values for Proxima Cen's movement and the detector parameters, assuming that Proxima was at the center of the detector. The position of Proxima was converted into sky coordinates using the Python routine \texttt{novas} \citep{2011AAS...21734414B}, which allows the sky coordinates of a star at a given epoch to be determined with an accuracy better than 1 mas. The stellar sky coordinates we adopted were the median of the RA and DEC derived for each epoch where the star was detected. We assumed that the proper motion and the parallax of these stars were negligible, using them as a fixed grid to readjust the detector parameters and Proxima Cen's movement. An attempt to include these parameters in the MCMC was made, but the fit did not converge. Note that since we assumed that the parallaxes of the background stars are negligible, the parallax found for Proxima is relative to the mean parallax of the background stars. To perform the MCMC analysis, we used the public Python code \texttt{emcee} \citep{2013PASP..125..306F}. We ran 130,000 steps, with 150 initial walkers. The result of the MCMC analysis for the plate scale and TN of each epoch as listed in Table~\ref{t:px}.

%An example of the results for one pair of epochs is shown in Figures~\ref{f:mcmc_scale}. In this case 10000 runs were performed, with 300 initial walkers, the final result is shown in Fig.~\ref{f:mcmc_scale}. After few runs the chains are stabilised around the final value. The pixel scale ratio, with respect to the nominal value of 12.25 mas/px, and TN correction applied for each epoch are listed in Table~\ref{t:px}.

%\begin{figure*}
%%%%%\vspace{8cm}
%\begin{center}
%\includegraphics[width=0.7\textwidth]{chainz_01_large}
%\caption{Position of each walker at each step in the chain of the MCMC. After few hundreds of steps the chains are stabilized around the final values.   }
%\label{f:chainz}
%\end{center}
%\end{figure*}

%\begin{figure*}
%%%%%\vspace{8cm}
%\begin{center}
%\includegraphics[width=0.7\textwidth]{post_jose_det}
%\caption{Results of the MCMC for one of the epochs. The parameters of the TN offset with respect to the nominal value (in degrees), the pixel scale in mas/pixel, and the centering in the two coordinates (in pixels) are shown.     }
%\label{f:mcmc_scale}
%\end{center}
%\end{figure*}

\begin{table}
 \centering
  \caption{Pixel scale and { True North (TN)} value for each epoch from the MCMC analysis.\label{t:px}}
  \begin{tabular}{c c c}
  \hline
    Epoch  & Pixel scale (mas/px) & TN (deg)  \\
         \hline
         2015-03-30  & 12.242$\pm 0.003$ &  -0.152$\pm 0.019$\\
         2016-01-18 & 12.243$\pm 0.004$ &  0.094$\pm 0.017$ \\
         2016-02-17 & 12.248$\pm 0.003$ &  0.195$\pm 0.015$\\
         2016-02-29 & 12.247$\pm 0.003$ &  -0.143$\pm 0.014$ \\
         2016-03-27 & 12.246$\pm 0.003$ &  -0.418$\pm 0.014$\\
         2016-04-15 & 12.243$\pm 0.003$ &  0.066$\pm 0.015$ \\
         2016-07-20 &  12.244$\pm 0.003$ &  0.003$^{\textrm{+}0.016}_{-0.015}$ \\
         2017-03-21  & 12.246$\pm 0.003$ &  0.147$\pm 0.019$ \\
         2017-06-15  & 12.240$\pm 0.004$ &  0.492$^{\textrm{+}0.026}_{-0.028}$   \\
         \hline
\end{tabular}
\end{table}

\begin{figure*}
%%%%%\vspace{8cm}
\begin{center}
\includegraphics[width=0.8\textwidth]{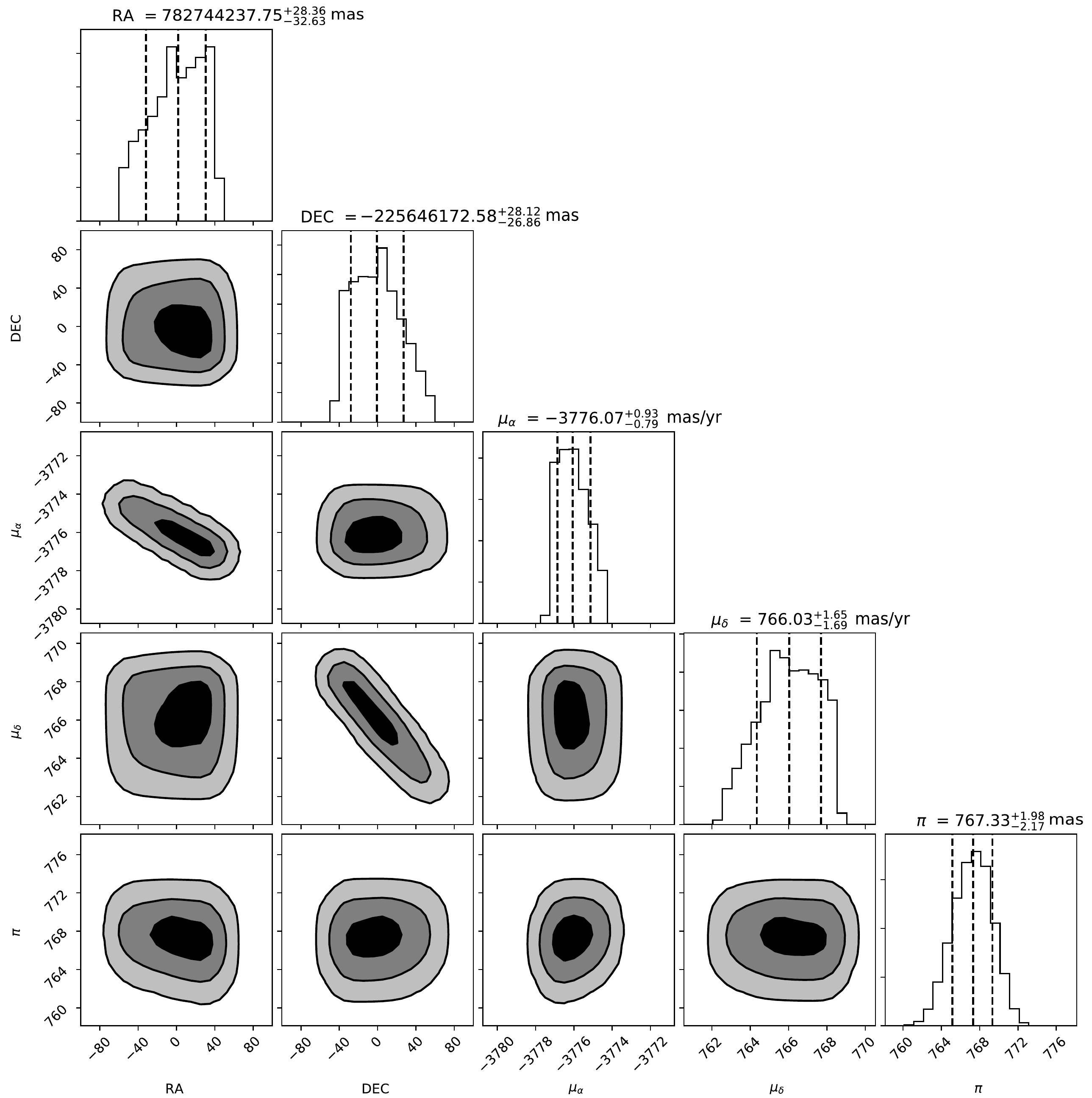}
\caption{{ Posterior probability distributions of the MCMC analysis of the relative movement of Proxima Cen with respect to the background stars.  The grey contours represent 68.3\%, 95.4\% and 99.7\% volume of the joint probability distribution, from dark to light grey, of the 2-D distributions. The vertical dashed lines represent the mean value and the uncertainties, which correspond to the 15.87th, 50th, 84.13th percentile levels of each distribution, as commonly done for multivariate MCMC results.}    }
\label{f:mcmc_prox}
\end{center}
\end{figure*}

The result of the MCMC analysis for the movement of Proxima Centauri on the sky is shown in Figure~\ref{f:mcmc_prox} and listed in Table~\ref{t:par_prox}. The value of the parallax found is in agreement with the measurements by \citet{1999AJ....118.1086B}, 768.7$\pm$0.3 mas, and \citet{2014AJ....148...91L}, 768.13$\pm$1.04 mas. This indicates that the correction between the parallax relative to the background stars and the absolute value is smaller than the error of the measurement.

{ To estimate the impact of neglecting PM and parallax of the background stars we performed another MCMC with this input: for each star we created synthetic measurements assuming as RA and Dec at the year 2016 the median value of the RA and Dec from our measurements. The parallax has been given randomly from a uniform distribution of distance between 1 and 3 kpc (we expect these distances for the background stars in the IRDIS FoV), while the PM has been calculated from a gaussian distribution of velocities with dispersion of 30 km/s, and random direction. From these synthetic data we derived again the parameters of the motion of Proxima Cen, and subsequently, the mass from the microlensing effect. The new value is $0.148^{\textrm{+}0.067}_{-0.055}$ \MSun and the fit has a reduced chi-square of 0.88. This measurement is consistent with the one that neglects PM and parallax of the background stars.}

%off = 150
%aux_first = res_mcmc[:,N:,:].reshape((-1,ndim))
%post_distr_fun = corner.corner(aux_first[:,0:5], quantiles=(0.1587, 0.5, 0.8413), levels=(1-np.exp(-0.5), 1-np.exp(-2), 1-np.exp(-4.5)), show_titles=True,title_fmt=".2f", title_args={"fontsize": 12},labels = [r"RA", r"DEC", r"$\mu_{\alpha}$",r"$\mu_{\delta}$", r"$\pi$"],range=[(ra_t-off, ra_t+off),(dec_t-off,dec_t+off),(mu_ra_t-5., mu_ra_t+5.), (mu_dec_t-5., mu_dec_t+5.), (pi_t-10., pi_t+10.)], \
%	 plot_datapoints=False, fill_contours=True, smooth=1.0, bins=20.)
%post_distr_fun.savefig('post_jose_prox.pdf')

\begin{table}
 \centering
  \caption{Parameters of Proxima Cen's motion on the sky with respect to the background stars from the MCMC analysis shown in Figure~\ref{f:mcmc_prox}. The values from Hipparcos \citep{2007A&A...474..653V} are also shown for a comparison.  \label{t:par_prox}}
  \begin{tabular}{c c c}
     & Measured relative motion & Hipparcos \\
  \hline
         \hline
         RA (J2000) & 14:29:42.949$^{\textrm{+0.029}}_{-0.033}$ & 14:29:42.949 \\     % 14.49526564\\
         DEC (J2000) & -62:40:46.172$^{\textrm{+0.028}}_{-0.027}$ & -62:40:46.163 \\                  % -62.67950506 \\
         $\mu_{RA}$   &  -3776.07$^{\textrm{+}0.93}_{-0.79}$ & -3775.75$\pm$1.63  mas/year\\
         $\mu_{Dec}$   &  766.06$^{\textrm{+}1.65}_{-1.69}$ & 765.54$\pm$2.01  mas/year\\
         $\pi$   & 767.33$^{\textrm{+}1.98}_{-2.17}$ & 771.64$\pm$2.60 mas \\
\hline
\end{tabular}
\end{table}

%_____________________________________________________________________

\section{Results}
\label{s:dis}

\subsection{Modelling of the microlensing event}
\label{s:micro}

After the determination of the accurate position of Proxima on the sky for each epoch, and the relative position of the background star Source 2, we measured the effect of the deflection of the light during the microlensing event. For that we have to consider that Source 2 itself has a { proper motion (PM)} and a parallax with respect to the other background stars. The parameters of the motion of the star in the sky are included as variables in the fit. In this fit we included also the WFC3/HST epoch of the 2013-03-31 to improve the baseline of the measurements of the relative positions between Proxima and Source 2. The reduction and analysis of this dataset is presented in Sahu et al., in prep. The position of the star was determined following the methods presented in Sec.~\ref{s:meas}.

The formula of the { angular Einstein radius} of a microlens is:

\begin{equation}
  \theta_{E}  $=$  \sqrt{\frac{4GM}{c^2d_{\pi}}},
  \label{form}
\end{equation}
where M is the mass of the lens, and 
\begin{equation}
\frac{1}{d_{\pi}} $=$ \frac{1}{D_L} - \frac{1}{D_{\star}}, 
\end{equation}
where D$_L$ and D$_{\star}$ are the distances from Earth to the lens and the background source, respectively. In this case we can approximate the distance of Source 2 to be infinite, as the source is located at $\sim$0.5 kpc (see discussion below). 

\begin{figure*}
%%%%%\vspace{8cm}
\begin{center}
\includegraphics[width=0.8\textwidth]{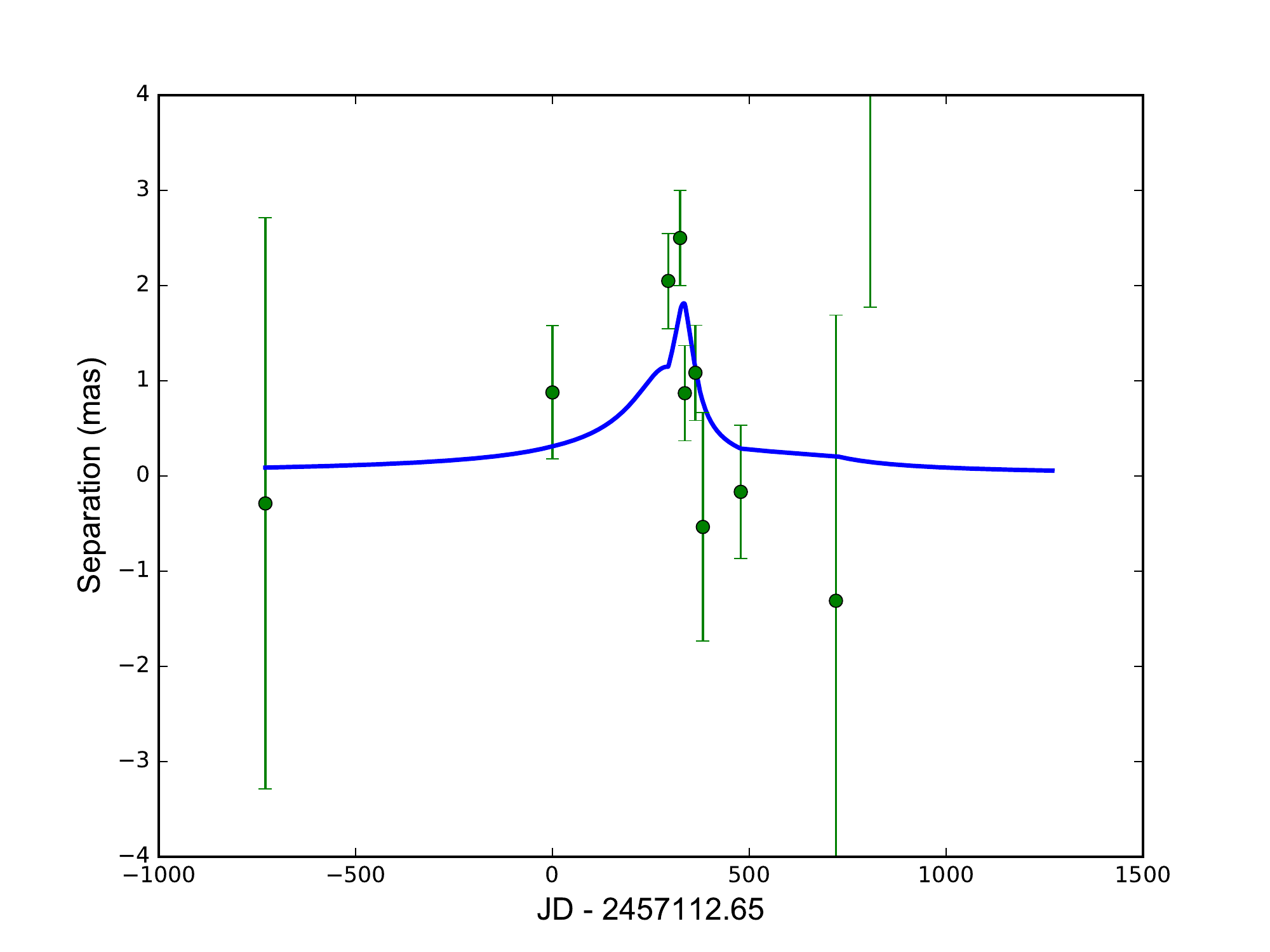}
\caption{{ Relative separation between Source 2 and Proxima Cen vs time in Julian days.} The solid blue line is the best fitting model { of the relative separation including the effect of the shifting due to the microlensing effect}. The first datapoint is WFC3/HST epoch of the 2013-03-31. The reference epoch is the first SPHERE epoch (2457112.65). The last epoch has a greater error bar as the star is just partially visible on the edge of the detector. Excluding this point does not change the result of the fit. Datapoints and model are presented in Table~\ref{t:amoeba}. }
\label{f:res_amoeba}
\end{center}
\end{figure*}

{The angular displacement of the images of a background source (with respect to its true position)} caused by the close passage of Proxima Cen can be written as:
\begin{equation}
  \theta_{\pm} $=$ 0.5 $($u \pm \sqrt{u^2 \textrm{+} 4}$)$\theta_E,
  \label{eq3}
\end{equation}
with $u$ = $\Delta \theta$/$\theta_E$, where $\Delta \theta$ is the angular separation between the lens and the undeflected Source 2, following the formalism of \citet{2014ApJ...782...89S}. What we measure in our data is the separation vector along the axis defined by the lens-source, $\theta_{\textrm{+}}$, which is shown in Fig.~\ref{f:res_amoeba}, while $\Delta \theta$ is calculated knowing the relative positions of Proxima and Source 2. From Eq.~\ref{eq3} we can then derive the angular Einstein radius $\theta_{E}$, and the mass of the Proxima, $M$. 

\begin{table*}
 \centering
  \caption{Results from the minimum $\chi$-square fit. { The relative measurements between Source 2 and Proxima Cen are compared to the expected quantities from the microlensing model.} The date, in Julian days, is relative to the first SPHERE epoch (2457112.65). All the units are expressed in mas except the position angle (PA) which is provided in degrees. \label{t:amoeba}}
  \begin{tabular}{l  l  l  l  l  l  l  l  l  l  l  l  l  l  }
     JD  & RA$_{mis}$  &  RA$_{mod}$  &   O-C  & Dec$_{mis}$ & Dec$_{mod}$  &  O-C &  Sep$_{mis}$ & Sep$_{mod}$   &  O-C &  PA$_{mis}$ & PA$_{mod}$ & O-C \\
\hline
         \hline
 -729.18 &-10522.31 & -10524.68 &   2.37  & 1925.77  & 1915.09 &  10.68 & 10697.08 & 10697.49  &  -0.41 &  280.37 &  280.31 &   0.06\\
    0.00 & -3001.82 &  -3001.07 &  -0.76  &  379.40  &  380.93 &  -1.53 &  3025.70 &  3025.15  &   0.56 &  277.20 &  277.23 &  -0.03\\
  294.20 &  -299.14 &   -299.21 &   0.07  & -764.09  & -763.09 &  -1.00 &   820.56 &   819.66  &   0.90 &  201.38 &  201.41 &  -0.03\\
  324.22 &    34.72 &     34.02 &   0.70  & -539.18  & -538.46 &  -0.72 &   540.30 &   539.53  &   0.77 &  176.32 &  176.39 &  -0.07\\
  336.23 &   221.02 &    221.20 &  -0.18  & -468.93  & -469.87 &   0.95 &   518.40 &   519.34  &  -0.93 &  154.76 &  154.79 &  -0.03\\
  363.12 &   730.88 &    730.35 &   0.54  & -386.63  & -387.75 &   1.12 &   826.85 &   826.89  &  -0.05 &  117.88 &  117.96 &  -0.09\\
  382.08 &  1140.03 &   1141.57 &  -1.54  & -402.02  & -401.59 &  -0.43 &  1208.84 &  1210.15  &  -1.31 &  109.42 &  109.38 &   0.04\\
  478.34 &  3026.99 &   3027.37 &  -0.38  &-1247.22  &-1247.48 &   0.26 &  3273.87 &  3274.32  &  -0.45 &  112.39 &  112.39 &  -0.00\\
  720.09 &  4332.93 &   4335.53 &  -2.60  &-1170.17  &-1166.40 &  -3.77 &  4488.16 &  4489.69  &  -1.53 &  105.11 &  105.06 &   0.06\\
  806.93 &  6222.03 &   6212.18 &   9.85  &-1593.26  &-1584.91 &  -8.35 &  6422.79 &  6411.17  &  11.61 &  104.36 &  104.31 &   0.05\\
\hline
\end{tabular}
\end{table*}

We performed a minimum $\chi$-squared fit with the routine \texttt{AMOEBA} to determine the unknown values of the PM and parallax of the star and the deflection due to the microlensing effect, $\theta_{\pm}$. { \texttt{AMOEBA} is an IDL function which performs a minimum $\chi$-squared fit using the downhill simplex method \citep{NelderMead65}.} The error bars assumed for each measurement of the relative position between Source 2 and Proxima Cen are chosen carefully following the methods presented in Sec.~\ref{s:budg}. In the fit, we also considered the effects of differential nutation, precession, and atmospheric diffraction (for SPHERE data). Even if these effects have a small contribution, they have to be taken into account when measuring the microlensing deflection.

%For the epochs where both IFS and IRDIS measurements were available, i.e. the critical epochs of February and March 2016, the errorbars were assumed to be 0.5 mas. For the last epoch of IRDIS the errorbar is greater as Source 2 was at the edge of the detector, partially visible.

The fit has a reduced $\chi$-square of 0.87, { there are 14 d.o.f., there are 20 fitted points (10 epochs and 2 coordinates), and there are 6 free parameters  (5 parameters of the motion of Source 2, the angular Einstein radius).} The best-fit results are listed in Table~\ref{t:amoeba} and the model of the deflection of the light due to the microlensing event is shown in Figure~\ref{f:res_amoeba}. A detailed table presenting the measurements and the model is presented in Table~\ref{t:amoeba}.

The contrast of Source 2 with respect to Proxima Cen is $\Delta$J=11.0$\pm$0.2 mag. The apparent magnitude of the object is thus mag$_J$=16.4$\pm$0.2. Given m$_V$=19.9 \citep[from][]{2014ApJ...782...89S}, the V-J color is equal to 3.5$\pm$0.3. This color is compatible with an M3$\pm$0.5 star \citep{1966ARA&A...4..193J} assuming no reddening. The absolute magnitude of such an object in the main sequence is M$_V$=11.1$\pm$1.0 mag \citep{2013ApJS..208....9P}, giving a distance modulus of 8.8$\pm$1.0 mag. The photometric distance of Source 2 is $\sim$ 0.58$_{-0.22}^{\textrm{+}0.32}$ kpc, or a parallax of 1.7$_{-0.6}^{\textrm{+}1.0}$ mas. This is slightly less than the parallax that we find with the microlensing model fit (3.25$\pm$0.39 mas). This value assumes a solar composition for Source 2, this may be the reason for the discrepancy between the photometric parallax and the parallax from the microlensing model. 

\begin{table}
 \centering
  \caption{Parameters of Source 2 on the sky with respect to the background stars and Einstein radius from the minimum $\chi$-squared analysis shown in Figure~\ref{f:mcmc_prox}. { Offsets in RA and Dec are relative to Proxima Cen. }\label{t:amoeba}}
  \begin{tabular}{c c }
  \hline
         \hline
         RA offset & -3003.92$\pm 0.35$  mas  \\     % 14.49526564\\
         DEC offset & 385.98$\pm 0.42$ mas \\                  % -62.67950506 \\
         $\mu_{RA}$   &  -1.65$\pm 0.61$ mas/year\\
         $\mu_{Dec}$   &  2.79$\pm 0.59$ mas/year\\
         $\pi$   & 3.25$\pm 0.39$  mas \\
         $\theta_E$   & 30.6$\pm 5.7$  mas \\
\hline
\end{tabular}
\end{table}

\subsection{Error budget}
\label{s:budg}

To estimate the error on the IRDIS measurement of each background star we employed the formula for the position presented in \citet{zurlo2014}, which provides a measurements error that depends on the S/N ratio of the point source. The errors on the movement of Proxima Cen are automatically generated by the MCMC procedure. The more complex estimation is the value of the error bars of the astrometric position of Source 2. For that, we assumed that the error bar related to the centering, TN, and plate scale are negligible after the correction of the detector parameters and the close distance of Source 2 to the center of the image. Accurate centering for these datasets was performed with the waffles present in each of the imaging epochs. This approach provided a centering accuracy of $\sim$0.3 mas. As presented in Table~\ref{t:px}, the errors on the plates cale and the TN direction are of the order of 0.003 mas/pixel and 0.015 deg, respectively.  At the 500 mas separation between Source 2 and Proxima Cen, these contribute to the error with the negligible value of 0.1 mas.

The uncertainty is therefore only estimated from the centering and the fitting procedure, i.e. the error bar is calculated as $\sigma$ = $\sqrt{\sigma^{2}_{cent} \textrm{+} \sigma^{2}_{FNP}}$. Where in this case the fitting procedure is fake negative planets (FNP) method routine applied on the IFS dataset. The errors calculated using the FNP are the standard deviation of the results of the fitting procedure while varying the modes of the PCA used for the reduction. Given the high S/N ratio of the source in the IFS datasets (above 20), and the fact that the errors on the calibration are negligible, we reached sub-milliarcsecond precision. For the epochs of February, March, and April, i.e. the critical epochs, we assumed an error bar of 0.5 mas. For the IRDIS measurements, the error was estimated using the same method described for the other background stars. For the last epoch, the error bar is larger (10 mas) because Source 2 is at the edge of the IRDIS detector and the star is only partially visible.     
  
\subsection{The mass of Proxima Centauri}
\label{s:star}

From the procedures described { we calculated $\theta_E$ and using the equation~\ref{form} of Sec.~\ref{s:micro}}, we obtained 0.150$^{\textrm{+}0.062}_{-0.051}$ \MSun for Proxima Centauri.  This is in agreement with the previous value assigned from luminosity-mass relations of $0.12\pm 0.02$~\MSun \citep{2015ApJ...804...64M}. We cannot further improve on this value with new SPHERE observations because Source 2 is no longer visible on the IRDIS detector when observing Proxima Cen. The precision of the astrometric position of this source is the highest ever reached with SPHERE, thanks to the exquisite quality of the data, and the calibration of the detector parameters with the large population of background stars in the FoV. Over the next few years, Proxima Cen will be followed-up to provide a better estimation of its movement on the sky. These data will be couple with observations from HST and Gaia to take advantage of future microlensing events. Unfortunately, no star as bright as Source 2 is expected to pass within 0\farcs5 of Proxima Cen in the next 20 years. Therefore only fainter stars will be usable for this purpose.

The minimum mass of the planet around Proxima Cen is Msini=1.27$^{\textrm{+}0.19}_{-0.17}$ \MEarth, assuming the host star mass of 0.1221 $\pm$ 0.0022 \MSun (\citep{2016Natur.536..437A}). If the microlensing derived mass of 0.150$^{\textrm{+}0.062}_{-0.051}$ \MSun is assumed for Proxima Cen, the minimum mass of the planet is estimated to be Msini=1.56$^{\textrm{+}0.064}_{-0.053}$ \MEarth.

\subsection{Constraints on a previously identified transit-like event}
\citet{2017RNAAS...1...49L} presented a candidate planetary transit around { Proxima Cen}. The depth of the candidate event was measured to be $\sim$0.005 mag in the I band. Given its phase, it could not be related to passage of the radial velocity detected planet (Proxima b) in front of the star. An eclipsing binary in the FWHM of the SPHERE IRDIS observations (5\arcsec) could have been the responsible for the signal detected transit like signal. However, there are no stars bright enough to mimic the transit in the IRDIS FoV. We calculate that a background eclipsing binary would need to be brighter than 13 mag in I band. Our SPHERE imaging data therefore rule out a background eclipsing binary as the source of the candidate transit and indicate that the event was produced by either another planet in the system \citep[but see][for a discussion]{2017RNAAS...1...49L}, some other astrophysical phenomenon (i.e. stellar activity), or noise.

%________________________________________________________________________

\section{Conclusions}
\label{s:conc}

Measurements of the parameters of Proxima Centauri, the closest star to our solar system, still suffer from significant uncertainties. In the past, its mass was indirectly determined using  low-mass star mass-luminosity relations, which suffer from significant systematic uncertainties. Last year, an approximately Earth mass planet in the habitable zone was discovered using the radial velocity technique. In 2016, a unique opportunity to directly measure the mass of Proxima Centauri occurred: the star, as seen from Earth, approached a background source with an impact parameter of 0\farcs5 and caused a microlensing event with Proxima Cen acting as the lens. In order to obtain precise astrometric measurement of the relative positions of the background source with respect to Proxima Cen and measure the deflection due to microlensing, we followed-up the star during a two-year monitoring program with SPHERE/IRDIFS. Nine epochs were obtained starting in March 2015.

We measured the deflection of the light of the background source caused by the close approach of Proxima Cen. The effect was revealed thanks to the exquisite sub-mas astrometric precision reached with SPHERE/IRDIFS. This was possible after careful recalibration of the detector parameters (true North, platescale, and centering) using the observations themselves as astrometric calibrators. The modelling of the { angular} Einstein radius of the microlensing effect gives a value of 0.150$^{\textrm{+}0.062}_{-0.051}$ \MSun for the mass of Proxima Cen. This is the first time that the gravitational mass of the closest star to our Solar system has been measured. Future observations of Proxima Cen may improve the precision on this estimate. Gaia measurements of the proper motion and parallax of Source 2 in the future may also refine the microlensing analysis and allow a more precise recalculation of Proxima Cen's mass.

%----------------------------------------------------------- S_vib

\section*{Acknowledgements}
We are grateful to the anonymous referee, after his/her revision the paper improved substantially. We are grateful to the SPHERE team and all the people at Paranal for the great effort during SPHERE GTO run. We are also grateful to Rodrigo F. D\'iaz for fruitful discussions. A.Z. acknowledges support from CONICYT through FONDECYT grant number 3170204. R.G., S.D. acknowledge support from the ``Progetti Premiali'' funding scheme of the Italian Ministry of Education, University, and Research. D.M. acknowledges support from the ESO-Government of Chile Joint Comittee
program ``Direct imaging and characterization of exoplanets''. We acknowledge support from the French National Research Agency (ANR) through the GUEPARD project grant ANR10-BLANC0504-01. H.A. acknowledges the financial support of the Swiss National Science Foundation through the NCCR PlanetS. SPHERE is an instrument designed and built by a consortium consisting of IPAG (Grenoble, France), MPIA (Heidelberg, Germany), LAM (Marseille, France), LESIA (Paris, France), Laboratoire Lagrange (Nice, France), INAF-- Osservatorio di Padova (Italy), Observatoire de Gen\`eve (Switzerland), ETH Zurich (Switzerland), NOVA (Netherlands), ONERA (France) and ASTRON (Netherlands), in collaboration with ESO. SPHERE was funded by ESO, with additional contributions from CNRS (France), MPIA (Germany), INAF (Italy), FINES (Switzerland) and NOVA (Netherlands). SPHERE also received funding from the European Commission Sixth and Seventh Framework Programmes as part of the Optical Infrared Coordination Network for Astronomy (OPTICON) under grant number RII3-Ct-2004-001566 for FP6 (2004-2008), grant number 226604 for FP7 (2009-2012) and grant number 312430 for FP7 (2013-2016). This work has made use of the SPHERE Data Centre,
jointly operated by OSUG/IPAG (Grenoble), PYTHEAS/LAM/CeSAM
(Marseille), OCA/Lagrange (Nice), and Observatoire de Paris/LESIA (Paris).

\bibliographystyle{mnras}
\bibliography{proxima_microlensing}

\bsp	% typesetting comment
\label{lastpage}
\end{document}